\RequirePackage{lineno}

\documentclass[aps,prd,twocolumn,showpacs,superscriptaddress,groupedaddress]{revtex4} 

\usepackage{graphicx}
\usepackage{dcolumn}
\usepackage{bm}

\begin{document}

\hspace{5.2in} \mbox{FERMILAB-PUB-16-038-E}

\title{Evidence for a {\boldmath  $B_s^0 \pi^{\pm}$} State}

\affiliation{LAFEX, Centro Brasileiro de Pesquisas F\'{i}sicas, Rio de Janeiro, RJ 22290, Brazil}
\affiliation{Universidade do Estado do Rio de Janeiro, Rio de Janeiro, RJ 20550, Brazil}
\affiliation{Universidade Federal do ABC, Santo Andr\'e, SP 09210, Brazil}
\affiliation{University of Science and Technology of China, Hefei 230026, People's Republic of China}
\affiliation{Universidad de los Andes, Bogot\'a, 111711, Colombia}
\affiliation{Charles University, Faculty of Mathematics and Physics, Center for Particle Physics, 116 36 Prague 1, Czech Republic}
\affiliation{Czech Technical University in Prague, 116 36 Prague 6, Czech Republic}
\affiliation{Institute of Physics, Academy of Sciences of the Czech Republic, 182 21 Prague, Czech Republic}
\affiliation{Universidad San Francisco de Quito, Quito, Ecuador}
\affiliation{LPC, Universit\'e Blaise Pascal, CNRS/IN2P3, Clermont, F-63178 Aubi\`ere Cedex, France}
\affiliation{LPSC, Universit\'e Joseph Fourier Grenoble 1, CNRS/IN2P3, Institut National Polytechnique de Grenoble, F-38026 Grenoble Cedex, France}
\affiliation{CPPM, Aix-Marseille Universit\'e, CNRS/IN2P3, F-13288 Marseille Cedex 09, France}
\affiliation{LAL, Univ. Paris-Sud, CNRS/IN2P3, Universit\'e Paris-Saclay, F-91898 Orsay Cedex, France}
\affiliation{LPNHE, Universit\'es Paris VI and VII, CNRS/IN2P3, F-75005 Paris, France}
\affiliation{CEA Saclay, Irfu, SPP, F-91191 Gif-Sur-Yvette Cedex, France}
\affiliation{IPHC, Universit\'e de Strasbourg, CNRS/IN2P3, F-67037 Strasbourg, France}
\affiliation{IPNL, Universit\'e Lyon 1, CNRS/IN2P3, F-69622 Villeurbanne Cedex, France and Universit\'e de Lyon, F-69361 Lyon CEDEX 07, France}
\affiliation{III. Physikalisches Institut A, RWTH Aachen University, 52056 Aachen, Germany}
\affiliation{Physikalisches Institut, Universit\"at Freiburg, 79085 Freiburg, Germany}
\affiliation{II. Physikalisches Institut, Georg-August-Universit\"at G\"ottingen, 37073 G\"ottingen, Germany}
\affiliation{Institut f\"ur Physik, Universit\"at Mainz, 55099 Mainz, Germany}
\affiliation{Ludwig-Maximilians-Universit\"at M\"unchen, 80539 M\"unchen, Germany}
\affiliation{Panjab University, Chandigarh 160014, India}
\affiliation{Delhi University, Delhi-110 007, India}
\affiliation{Tata Institute of Fundamental Research, Mumbai-400 005, India}
\affiliation{University College Dublin, Dublin 4, Ireland}
\affiliation{Korea Detector Laboratory, Korea University, Seoul, 02841, Korea}
\affiliation{CINVESTAV, Mexico City 07360, Mexico}
\affiliation{Nikhef, Science Park, 1098 XG Amsterdam, the Netherlands}
\affiliation{Radboud University Nijmegen, 6525 AJ Nijmegen, the Netherlands}
\affiliation{Joint Institute for Nuclear Research, Dubna 141980, Russia}
\affiliation{Institute for Theoretical and Experimental Physics, Moscow 117259, Russia}
\affiliation{Moscow State University, Moscow 119991, Russia}
\affiliation{Institute for High Energy Physics, Protvino, Moscow region 142281, Russia}
\affiliation{Petersburg Nuclear Physics Institute, St. Petersburg 188300, Russia}
\affiliation{Instituci\'{o} Catalana de Recerca i Estudis Avan\c{c}ats (ICREA) and Institut de F\'{i}sica d'Altes Energies (IFAE), 08193 Bellaterra (Barcelona), Spain}
\affiliation{Uppsala University, 751 05 Uppsala, Sweden}
\affiliation{Taras Shevchenko National University of Kyiv, Kiev, 01601, Ukaine}
\affiliation{Lancaster University, Lancaster LA1 4YB, United Kingdom}
\affiliation{Imperial College London, London SW7 2AZ, United Kingdom}
\affiliation{The University of Manchester, Manchester M13 9PL, United Kingdom}
\affiliation{University of Arizona, Tucson, Arizona 85721, USA}
\affiliation{University of California Riverside, Riverside, California 92521, USA}
\affiliation{Florida State University, Tallahassee, Florida 32306, USA}
\affiliation{Fermi National Accelerator Laboratory, Batavia, Illinois 60510, USA}
\affiliation{University of Illinois at Chicago, Chicago, Illinois 60607, USA}
\affiliation{Northern Illinois University, DeKalb, Illinois 60115, USA}
\affiliation{Northwestern University, Evanston, Illinois 60208, USA}
\affiliation{Indiana University, Bloomington, Indiana 47405, USA}
\affiliation{Purdue University Calumet, Hammond, Indiana 46323, USA}
\affiliation{University of Notre Dame, Notre Dame, Indiana 46556, USA}
\affiliation{Iowa State University, Ames, Iowa 50011, USA}
\affiliation{University of Kansas, Lawrence, Kansas 66045, USA}
\affiliation{Louisiana Tech University, Ruston, Louisiana 71272, USA}
\affiliation{Northeastern University, Boston, Massachusetts 02115, USA}
\affiliation{University of Michigan, Ann Arbor, Michigan 48109, USA}
\affiliation{Michigan State University, East Lansing, Michigan 48824, USA}
\affiliation{University of Mississippi, University, Mississippi 38677, USA}
\affiliation{University of Nebraska, Lincoln, Nebraska 68588, USA}
\affiliation{Rutgers University, Piscataway, New Jersey 08855, USA}
\affiliation{Princeton University, Princeton, New Jersey 08544, USA}
\affiliation{State University of New York, Buffalo, New York 14260, USA}
\affiliation{University of Rochester, Rochester, New York 14627, USA}
\affiliation{State University of New York, Stony Brook, New York 11794, USA}
\affiliation{Brookhaven National Laboratory, Upton, New York 11973, USA}
\affiliation{Langston University, Langston, Oklahoma 73050, USA}
\affiliation{University of Oklahoma, Norman, Oklahoma 73019, USA}
\affiliation{Oklahoma State University, Stillwater, Oklahoma 74078, USA}
\affiliation{Oregon State University, Corvallis, Oregon 97331, USA}
\affiliation{Brown University, Providence, Rhode Island 02912, USA}
\affiliation{University of Texas, Arlington, Texas 76019, USA}
\affiliation{Southern Methodist University, Dallas, Texas 75275, USA}
\affiliation{Rice University, Houston, Texas 77005, USA}
\affiliation{University of Virginia, Charlottesville, Virginia 22904, USA}
\affiliation{University of Washington, Seattle, Washington 98195, USA}
\author{V.M.~Abazov} \affiliation{Joint Institute for Nuclear Research, Dubna 141980, Russia}
\author{B.~Abbott} \affiliation{University of Oklahoma, Norman, Oklahoma 73019, USA}
\author{B.S.~Acharya} \affiliation{Tata Institute of Fundamental Research, Mumbai-400 005, India}
\author{M.~Adams} \affiliation{University of Illinois at Chicago, Chicago, Illinois 60607, USA}
\author{T.~Adams} \affiliation{Florida State University, Tallahassee, Florida 32306, USA}
\author{J.P.~Agnew} \affiliation{The University of Manchester, Manchester M13 9PL, United Kingdom}
\author{G.D.~Alexeev} \affiliation{Joint Institute for Nuclear Research, Dubna 141980, Russia}
\author{G.~Alkhazov} \affiliation{Petersburg Nuclear Physics Institute, St. Petersburg 188300, Russia}
\author{A.~Alton$^{a}$} \affiliation{University of Michigan, Ann Arbor, Michigan 48109, USA}
\author{A.~Askew} \affiliation{Florida State University, Tallahassee, Florida 32306, USA}
% ...
%
%
\collaboration{The D0 Collaboration\footnote{with visitors from
$^{a}$Augustana College, Sioux Falls, SD 57197, USA,
$^{b}$The University of Liverpool, Liverpool L69 3BX, UK,
$^{c}$Deutshes Elektronen-Synchrotron (DESY), Notkestrasse 85, Germany,
$^{d}$CONACyT, M-03940 Mexico City, Mexico,
$^{e}$SLAC, Menlo Park, CA 94025, USA,
$^{f}$University College London, London WC1E 6BT, UK,
$^{g}$Centro de Investigacion en Computacion - IPN, CP 07738 Mexico City, Mexico,
$^{h}$Universidade Estadual Paulista, S\~ao Paulo, SP 01140, Brazil,
$^{i}$Karlsruher Institut f\"ur Technologie (KIT) - Steinbuch Centre for Computing (SCC),
D-76128 Karlsruhe, Germany,
$^{j}$Office of Science, U.S. Department of Energy, Washington, D.C. 20585, USA,
$^{k}$American Association for the Advancement of Science, Washington, D.C. 20005, USA,
$^{l}$Kiev Institute for Nuclear Research (KINR), Kyiv 03680, Ukraine,
$^{m}$University of Maryland, College Park, MD 20742, USA,
$^{n}$European Orgnaization for Nuclear Research (CERN), CH-1211 Geneva, Switzerland,
$^{o}$Purdue University, West Lafayette, IN 47907, USA,
and
$^{p}$Lebedev Physical Institute of the Russian Academy of Sciences, Moscow, 119991, Russia.
$^{\ddag}$Deceased.
}} \noaffiliation
\vskip 0.25cm

\date{\today}
           
\begin{abstract}

We report evidence for a narrow structure, $X(5568)$, 
in the decay sequence $X(5568) \rightarrow B_s^0 \pi^{\pm}$,
 $B_s^0 \rightarrow J/\psi \phi$, $J/\psi \rightarrow \mu^+ \mu^-$, $\phi \rightarrow K^+K^-$.
This is evidence for the first instance of a hadronic state
with valence quarks of four different flavors.
The mass and natural width of this state are measured to be
$m = 5567.8 \pm  2.9  {\rm \thinspace (stat)} ^{+0.9}_{-1.9}  {\rm \thinspace (syst)}$~MeV/$c^2$
and 
$\Gamma = 21.9 \pm 6.4  {\rm \thinspace (stat)}   ^{+5.0}_{-2.5} {\rm \thinspace (syst)} $~MeV/$c^2$.
If the decay is $X(5568) \rightarrow 
B_s^* \pi^{\pm} \rightarrow B_s^0 \gamma \pi^{\pm}$ with an unseen $\gamma$, $m(X(5568))$ will be 
shifted up by $m(B_s^*) - m(B_s^0) \sim 49$~MeV/$c^2$.
This measurement is based on
$10.4~\rm{fb^{-1}}$ of $p \overline p $ collision data at $\sqrt{s}\,=$ 1.96 TeV
collected by the D0 experiment at the Fermilab Tevatron collider.

\end{abstract}


\maketitle


\newpage

During the last few years several resonant states that cannot be
conventional quark-antiquark mesons or three-quark baryons
have been observed~\cite{bela,belb,cdfa,belc,besa,beld,besb,lhcba}. Taking into account the
decay modes and charges of these states, they may be 
interpreted as four-quark or five-quark states. These 
states  have one common feature:
they consist of a combination of heavy and light quarks.
These discoveries open up a new era of multiquark hadron spectroscopy.
Various combinations of heavy and light mesons may be tested.
One such system is the combination of the heavy $B_s^0$ or $\overline B_s^0$ meson and 
the light $\pi^{\pm}$ meson. 
Such systems  are composed of two quarks and two antiquarks of four different flavors:
$b, s, u, d$, which
might be a tightly bound diquark antidiquark pair
such as $[bu][\overline d \overline s]$, $[bd][\overline s \overline u]$,
$[su][\overline b \overline d]$, or $[sd][\overline b \overline u]$,
or a ``molecule'' of the loosely bound $B$ and $K$ mesons.
This Letter presents a study of 
the $B_s^0 \pi^{\pm}$ invariant mass spectrum using a data 
sample of 10.4 fb$^{-1}$ collected with the
D0 detector at the Fermilab Tevatron collider.

The D0 detector consists of a central tracking system, calorimeters, and
muon detectors~\cite{Abazov2006463}. The central
tracking system comprises  a silicon microstrip tracker (SMT) and a central
fiber tracker (CFT), both located inside a 1.9~T superconducting solenoidal
magnet.  The tracking system is designed to optimize tracking and vertexing
for pseudorapidities $|\eta|<3$,
where  $\eta = -\ln[\tan(\theta/2)]$, and  $\theta$ is the 
polar angle with respect to the proton beam direction.
  The SMT can reconstruct the $p\overline{p}$ interaction vertex (primary vertex) 
for interactions   with at least three tracks with a precision
of 40~$\mu$m in the plane transverse to the beam direction.
  The muon detector, positioned outside the calorimeter, consists of a central muon system covering the
 pseudorapidity region  $|\eta|<1$ and a forward muon system covering the pseudorapidity region
  $1<|\eta|<2$. 
Both central and forward systems consist of a layer of drift  tubes
and scintillators inside 1.8~T iron toroidal magnets with two similar layers outside the toroids.

Events used in this analysis are collected with both single muon  and dimuon triggers.
Single muon triggers require a coincidence of signals in trigger elements inside and outside
the toroidal magnets.
Dimuon triggers in the central rapidity region  require at least one muon to penetrate the toroid.
In the forward region, both muons are required to penetrate the toroid.

Candidate events are required to include a pair of oppositely charged muons 
both with $p_T>1.5$~GeV/$c$ in the invariant mass range
 $2.92<m(\mu^+ \mu^-)<3.25$~GeV/$c^2$, consistent with $J/\psi$ decay, accompanied by two additional  particles of
 opposite charge assumed to be kaons, each with $p_T>0.7$~GeV/$c$,  with an invariant mass of 
  $1.012 <m(K^+K^-)<1.030$~GeV/$c^2$, consistent with $\phi$ decay, and a third charged particle with  $p_T>0.5$~GeV/$c$
assumed to be a pion.

  In the event selection, both muons are required to be
detected  in the muon chambers inside the toroidal magnet, and at least one
of the muons is required to be also detected
outside the iron toroid.
Each muon candidate~\cite{muid} is required to match a track found in the central tracking system, and
each of the five final-state tracks is required to have at least one SMT hit and at least one CFT hit.
The dimuon invariant mass is constrained  to the world-average   $J/\psi$ mass~\cite{pdg2014}, and 
 the four tracks forming a $J/\psi \phi$ candidate  are required to satisfy a fit to a 
 common vertex that is displaced from 
the primary vertex in the plane perpendicular to the beam direction
by at least 3 times the standard deviation of the measurement  uncertainty.
The pion candidate is required to be consistent with originating from the primary $p \overline p$ collision vertex.

\begin{figure}
\includegraphics[scale=0.40]{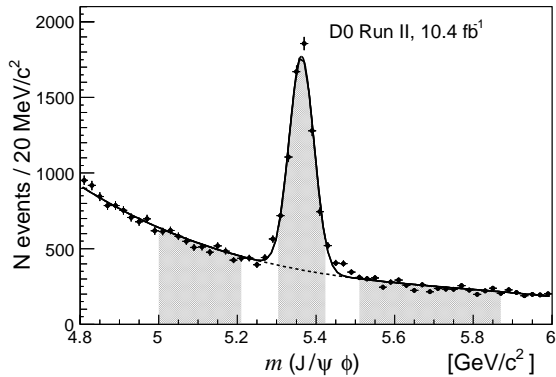}
\caption{\label{fig:epsart} Invariant mass distribution of $J/\psi \phi$ candidates.
The signal region and two sideband regions are indicated. The solid curve 
presents the fit results to the function, modeled by a sum of a
third-order polynomial to describe the combinatorial background and 
a Gaussian to describe the $B_s^0$ signal. The dotted curve shows
the combinatorial background.}
 \label{fig:mbs}
\end{figure}

To form a $B_s^0 \pi^{\pm}$ combination, we select the $B_s^0$ 
candidates in the mass range $5.303< m(J/\psi \phi) <5.423$~GeV/$c^2$,  
corresponding to an interval of  $\pm 2$ standard deviations around the mean value of the reconstructed $B_s^0$ mass. 
The $m\,(J/\psi \phi)$ distribution is shown in Fig.~\ref{fig:mbs}.
The fit, including a third-order polynomial describing the combinatorial
background and a Gaussian function describing the signal,  yields the Gaussian
signal parameters $m(B_s^0)$ = 5363.3 $\pm$ 0.6 MeV/$c^2$, $\sigma (B_s^0)$ = 31.6 $\pm$ 0.6 MeV/$c^2$
and the number of signal events $N_{\rm ev}$ = 5582 $\pm$ 100. 
To improve the resolution of the invariant mass of the $B_s^0 \pi^{\pm}$
system and to remove the measured $B_s^0$ mass bias, we define it as 
$m(B_s^0 \pi^{\pm}) = m (J/\psi \phi\, \pi^{\pm}) - m (J/\psi \phi) + 5.3667 $ GeV/$c^2$,
where $m(J/\psi)$ is not constrained to the nominal value.
We study events as a function of mass in the range $5.5<m\,(J/\psi \phi)<5.9$~GeV/$c^2$.

Background in the $B_s^0 \pi^{\pm}$ invariant mass spectrum
results from  random combinations of selected $B_s^0$ candidates with
low momentum  charged particles coming mostly from the primary vertex.
To suppress background the $B_s^0 \pi^{\pm}$ system is required to have $p_T>10$~GeV/$c$.
To further reduce background, we impose a limit on the difference between the directions of the
 $B_s^0$ candidate and the pion to be 
$\Delta R = \sqrt{\Delta\eta^2+\Delta\phi^2}<0.3$, where $\eta$ is the pseudorapidity
and $\phi$ is the azimuthal angle. In addition to increasing the signal-to-background ratio
this ``cone cut''   limits  backgrounds  that  are not included in available
simulations.

The $B_s^0$ candidates include  genuine $B_s^0$
mesons and  the combinatorial background under the $B_s^0$ signal, as seen in Fig.~\ref{fig:mbs}.
The $B_s^0 \pi^{\pm}$ background with a real $B_s^0$ meson is modeled
using a Monte Carlo (MC) simulation~\cite{Pythia} of events containing a $B_s^0$ meson and additional pions
tuned to reproduce the $B_s^0$ transverse momentum distribution in data.

The background with a false $B_s^0$ meson is modeled using the
sideband events obtained from data.
The chosen sideband regions 5.0 $< m(J/\psi \phi) <$ 5.21 GeV/$c^2$
and 5.51 $< m(J/\psi \phi) <$ 5.87 GeV/$c^2$ are indicated in Fig.~\ref{fig:mbs}.
The sidebands are separated by $\sim$5$\sigma$ from the $B_s^0$ nominal mass.
The left and right sideband ranges are chosen 
to provide a large event sample and to have an average mass of  $m(B_s^0)$.

The two background components are found to have similar shapes~\cite{epaps}.
The fraction of the real $B_s^0$ events in the signal region is
obtained from the fit to  the $B_s^0$ meson in the $m(J/\psi \phi)$  distribution 
and is found to be (70.9 $\pm$ 0.6)$\%$.
MC  events and the sideband events are mixed in this proportion
to obtain the combined  background  that includes  pions from both
sources. The event selection results in
 pions that mainly come from the primary vertex, although pions originating 
 from heavy flavor decays are also present
in the sample.

Multiple entries for a single event may occur when more than one
pion candidate passes the event selection and they are retained in the sample.
The rate of duplicate entries
in the mass range $5.5<m(B_s^0 \pi^{\pm})  <5.6$~GeV/$c^2$  
($\sim$5$\%$)
is lower than for masses above 5.7 GeV/$c^2$   ($\sim$8\%).

\begin{figure}
\includegraphics[scale=0.40]{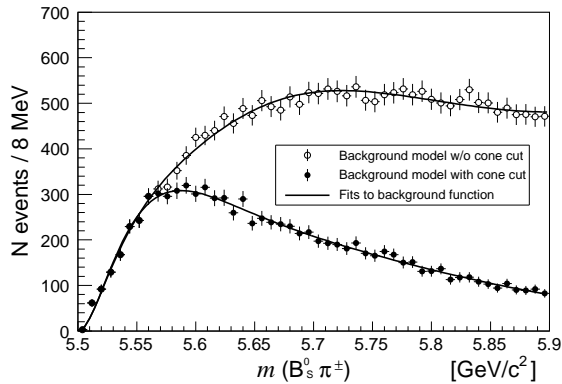}

\caption{\label{fig:noconebkg}  
The combined  background for the $m(B_s^0 \pi^{\pm})$ distribution described in the text 
and the fit to that distribution
with the $\Delta R < 0.3$ cone cut and without the cone cut.}
\label{fig:mbkgcomp}
\end{figure}

The combined background is
modeled by a function of the parameter $m_0 = m_{B\pi}  - \Delta$, 
where $m_{B\pi} \equiv m(B_s^0 \pi^{\pm})$ and $\Delta = 5.5$  GeV/$c^2$,
 of the form

\begin{equation}
F_{\rm {bgr}} (m_0) = P_{4 (C1 = 0)} \, \exp \left( P_2 \right).
\end{equation}
\noindent
Here, $P_{4 (C1=0)}$ and $P_2$ are fourth- and second-order polynomials, and the linear term of the first polynomial is
set to zero.
This empirical function gives a good description of the combined backgrounds,
as seen in Fig.~\ref{fig:mbkgcomp}.

\begin{figure}
\includegraphics[scale=0.40]{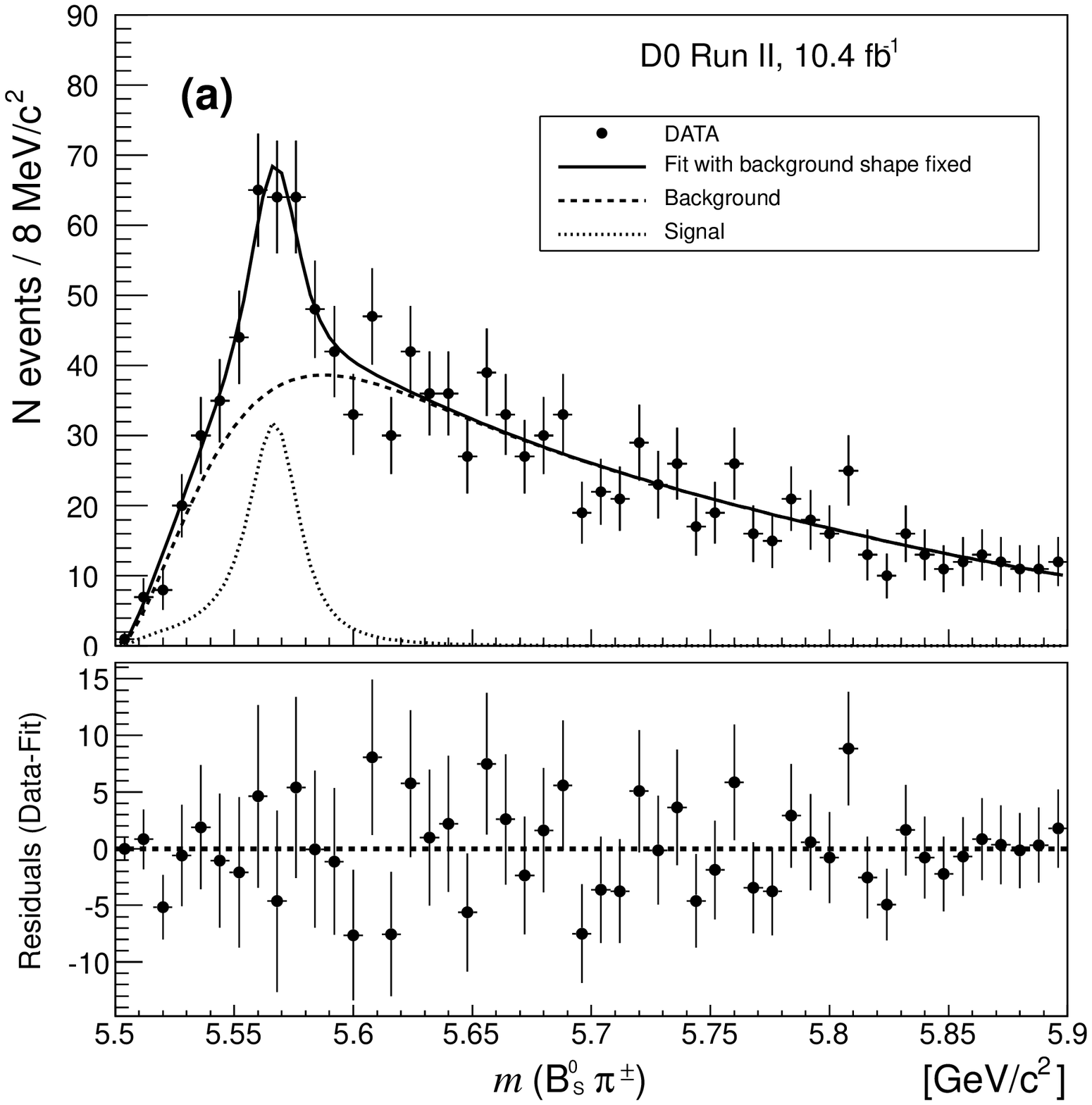}
\includegraphics[scale=0.40]{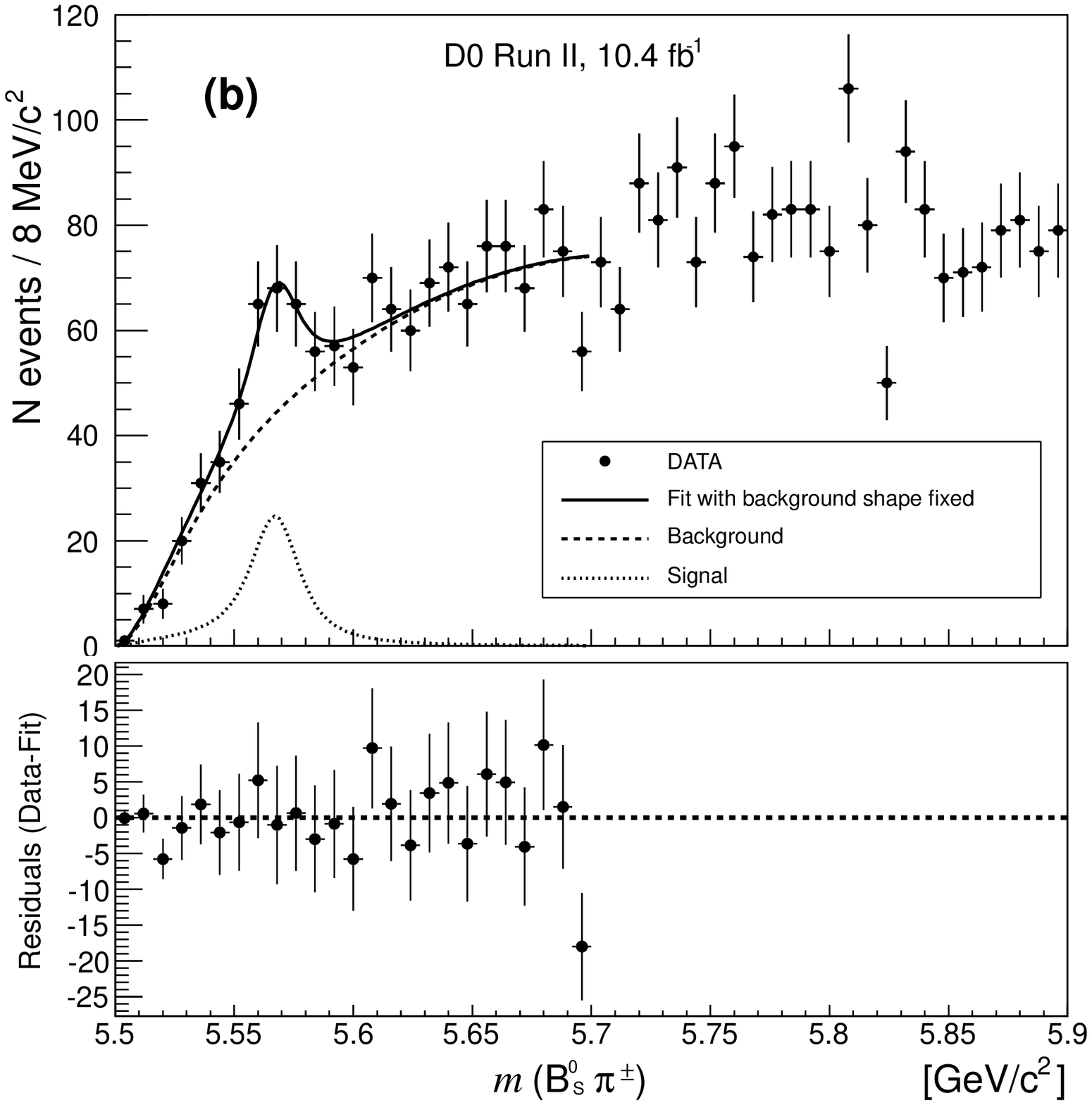}
\caption{\label{fig:epsart} The $m(B_s^0 \pi^{\pm})$ distribution
together with the background distribution and the fit results (a) after
applying the $\Delta R < 0.3$ cone cut and (b) without the cone cut. }
\label{fig:mbspidr}
\end{figure}

The $B_s^0 \pi^{\pm}$ invariant mass spectrum
is shown in Fig.~\ref{fig:mbspidr}(a) with the cone cut
and (b) without the cone cut. An enhancement is seen
near 5.57 GeV/$c^2$. To extract the signal parameters,
the distributions are fitted with a function $F$ [Eq. (2)]
that includes two terms: the background term  $F_{\rm bgr}(m_{B\pi})$
with fixed shape parameters
as in Fig.~\ref{fig:mbkgcomp} and
the signal term $F_{\rm sig}(m_{B\pi}, M_X, \Gamma{_X})$, modeled
by a relativistic Breit-Wigner function convolved with a Gaussian 
detector resolution function and with the mass-dependent efficiency of
the cone cut~\cite{epaps}. Here $M_X$ and $\Gamma_X$ are the mass and the natural 
width of the resonance.
The Gaussian width parameter  $\sigma_{\rm res} = 3.8$~MeV/$c^2$ is taken
from simulations.

The fit function has the form

\begin{equation}
   F = f_{\rm sig} \, F_{\rm sig}(m_{B\pi}, M_X, \Gamma{_X}) + f_{\rm bgr} \, F_{\rm bgr} (m_{B\pi}) ,
\end{equation}
\noindent
where $f_{\rm sig}$ and $f_{\rm bgr}$ are normalization factors.

We use the Breit-Wigner parametrization appropriate for an $S$-wave
two-body decay near threshold:

\begin{equation}
BW(m_{B\pi}) \propto \frac{ M_X^2 \Gamma (m_{B\pi})}{(M_X^2 - m_{B\pi}^2)^2 + M_X^2 \Gamma{^2} (m_{B\pi})}. 
\end{equation}
The mass-dependent width 
$\Gamma (m_{B\pi})=\Gamma_X \cdot (q_1 / q_0) $ is proportional to the natural width $\Gamma_X$, 
where $q_1$ and $q_0$ are three-vector momenta of the $B_s^0$
meson in the rest frame of the $B_s^0 \pi^{\pm}$ system at the invariant mass
equal  to $m_{B\pi}$ and $M_X$, respectively.

In the fit shown in Fig.~\ref{fig:mbspidr}(a),
the normalization parameters $f_{\rm sig}$ and $f_{\rm bgr}$
and the Breit-Wigner parameters $M_X$ and $\Gamma{_X}$ are allowed
to vary.
The fit yields the mass and width of
$M_X = 5567.8 \pm 2.9$~MeV/$c^2$,
$\Gamma{_X} = 21.9 \pm 6.4$~MeV/$c^2$, and the number of signal events of
$N = 133 \pm 31$. As the measured width is significantly larger than the experimental mass resolution,
 we infer that  $X(5568) \rightarrow  B_s^0 \pi^{\pm}$ is a strong decay.
The statistical significance of the signal is defined as 
$\sqrt{-2\, {\rm ln} ({\cal{L}}_0 /{\cal{L}}_{\rm max}) }$,
where ${\cal{L}}_{\rm max}$ and ${\cal{L}}_0$   are likelihood values at the
best-fit signal yield and the signal yield fixed to zero.
The obtained local statistical 
significance is 6.6$\sigma$ for the given mass and width values.
With the look-elsewhere effect ~\cite{lee} taken into account, 
the global statistical significance is 6.1$\sigma$.
The search window is taken as the interval between the $B_s^0 \pi^{\pm}$ 
threshold (5506 MeV/$c^2$) and the $B_d^0 K^{\pm}$ mass threshold (5774 MeV/$c^2$).

We also extract the signal from the $m(B_s^0 \pi^\pm)$ distribution 
without the $\Delta R$ cone cut, fixing the mass and natural width of 
the signal and the background mass shape to their default values.  We 
see a tendency for data to exceed background for
$m(B_s^0 \pi^\pm) > M_X$~\cite{epaps}.   We perform a fit in the 
restricted range $m(B_s^0 \pi^\pm) < 5.7$ GeV/$c^2$ [Fig. 3(b)] and find the 
fitted number of signal events to be $106 \pm 23$, with a corresponding
local statistical significance of 4.8$\sigma$.
The difference in 
yields with and without the cone cut is not fully explained by 
statistical fluctuations. 
In a subsidiary study we used empirical functions~\cite{X4140} for the background fitted to
the sidebands in data below the $X(5568)$ region 
and above the signal region up to 5.9 GeV/$c^2$
and found signal yields that are greater than those with
the default background function and comparable to
or greater than that found in the cone cut analysis.
These results confirm that using a background function that agrees with data for masses above 5.7 GeV 
can increase the fitted signal yield above that obtained using the default background model.
Additional background 
processes not present in our MC calculations such as $B_c \rightarrow B_s\thinspace n\pi$ with 
$n>1$, or other new states at higher mass, would thus have the effect of 
reducing the $X(5568)$ yield for the no-cone cut case.

\begin{figure}
\includegraphics[scale=0.40]{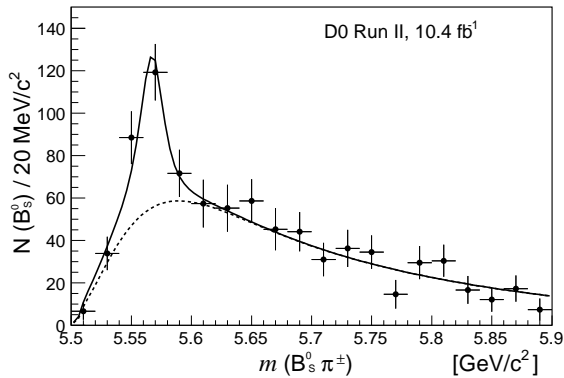}
\caption{The $m(B_s^0 \pi^{\pm})$ distribution
resulting from fits of the $B_s^0$ signal to  $m(J/\psi \phi$)
in twenty  mass intervals of $(B_s^0 \pi^{\pm})$  candidates 
described in  the text.
The solid line shows the result of the fit. The dashed line shows the contribution
of events due to the $B_s^0$ background from simulations.
There is no non-$B_s^0$ background. 
}
\label{fig:mbins}
\end{figure}

As a cross-check, we extract a pure $B_s^0 \pi^{\pm}$ signal by performing
fits of the number of $B_s^0$ events in the  $J/\psi \phi$ mass distribution in 20 MeV/$c^2$ intervals 
in the range $5.5<m(B_s^0\pi^{\pm})< 5.9$~MeV/$c^2$.
Results of those fits are  shown in Fig.~\ref{fig:mbins}.
A fit to the dependence of resulting $B_s^0$ yields on  $m(B_s^0 \pi^{\pm})$,
with the mass and natural width fixed to the previously obtained values, gives $118 \pm 22$ signal events. This result confirms that the observed signal is due to $B_s^0 \pi^{\pm}$ candidates
with genuine $B_s^0$ mesons and thus eliminates the possibility of non-$B_s^0$ processes
mimicking the signal.

\begin{table*}
\caption{\label{tab:table1} Systematic uncertainties for the observed
$X(5568)$ state mass, natural width and number of events. }
\begin{ruledtabular}
\begin{tabular}{lccc}
Source & Mass, MeV/$c^2$ & Width, MeV/$c^2$  & Rate, $\%$ \\
\hline
 {\it Background shape} &  &  &  \\
~~ MC samples with soft or hard $B_s^0$ & +0.2 \ ;\ $-0.6$  & +2.6 \ ;\ $-0.0$ & +8.2 \ ;\  $-0.0$ \\
~~ Sideband mass ranges  & +0.2 \ ;\ $-0.1$ & +0.7 \ ;\ $-1.7$  & +1.6 \ ;\ $-9.3$ \\
~~ Sideband mass calculation method  & +0.1 \ ;\ $-0.0$ & +0.0 \ ;\ $-0.4$ & +0.0 \ ;\ $-1.3$ \\
~~ MC to sideband events ratio  & +0.1 \ ;\ $-0.1$ & +0.5 \ ;\ $-0.6$ & +2.8 \ ;\ $-3.1$ \\
~~ Background function used  & +0.5 \ ;\ $-0.5$ & +0.1 \ ;\ $-0.0$ & +0.2 \ ;\ $-1.1$ \\
~~ $B_s^0$ mass scale, MC and data & +0.1 \ ;\ $-0.1$ & +0.7 \ ;\ $-0.6$ & +3.4 \ ;\ $-3.6$ \\
\hline
{\it Signal shape}  &  &  &  \\
~~ Detector resolution  & +0.1 \ ;\ $-0.1$ & +1.5 \ ;\ $-1.5$ & +2.1 \ ;\ $-1.7$ \\
~~ Non-relativistic BW  & +0.0 \ ;\ $-1.1$ & +0.3 \ ;\ $-0.0$ & +3.1 \ ;\ $-0.9$ \\
~~ {\it P}-wave BW  & +0.0 \ ;\ $-0.6$ & +3.1 \ ;\ $-0.0$ & +3.8 \ ;\ $-0.0$ \\
\hline
{\it Other}  &  &  &  \\
~~ Binning  & +0.6 \ ;\ $-1.1$ & +2.3 \ ;\ $-0.0$ & +3.5 \ ;\ $-3.3$ \\
\hline 
Total & +0.9 \ ;\ $-1.9$ & +5.0 \ ;\ $-2.5$ & +11.4 \ ;\ $-11.2$ \\
\end{tabular}
\end{ruledtabular}
\end{table*}

We obtain the systematic uncertainties  for the measured values of
the $X(5568)$ state mass, natural width, and the number of events.
The dominant uncertainties are due to
the background  and signal shapes.
We evaluate the systematic uncertainties due to
the background shape by (i) using different models of bottom pair production
in generating the $B_s^0$ MC samples,
 (ii) varying 
the sideband mass intervals, (iii) changing  the way the $B_s^0$ mass constraint
is applied in the calculation of $m(B_s^0 \pi^{\pm})$ 
for the sideband events by replacing the mass difference defined in the text by the kinetic
energy obtained by forcing $m(J/\psi \phi)$ to the world-average $B_s^0$ mass,
  (iv) changing the ratio of the MC to
the sideband events within 1$\sigma$, (v) using different background
functions by replacing the fourth-order polynomial in Eq. (1) with a
third- or fifth-order polynomial
or replacing the second-order polynomial in the exponential with the first-
or third-order polynomial,
 and (vi) varying
the nominal $B_s^0$ mass within $\pm 1$ MeV/$c^2$ in the background samples,
both for the sideband data and  simulated events.
The systematic uncertainties due to the signal shape are evaluated
by (i) varying the detector resolution within $\pm$1 MeV/$c^2$ around the mean value,
(ii) using a nonrelativistic Breit-Wigner function, and (iii) using a $P$-wave
relativistic Breit-Wigner function.

Additionally, we estimate the systematic uncertainties due to the  binning
by changing the bin size to 5 MeV/$c^2$,  
and to 10 MeV/$c^2$ instead of 8 MeV/$c^2$,
and shifting the lower edge  of
the mass scale by 1/3, 1/2, and 2/3 of the bin size. 
All systematic uncertainty sources are summarized in Table 1.
The uncertainties are added in quadrature separately for positive and 
negative values to obtain the total systematic uncertainties for each 
measured parameter and are treated as nuisance parameters to construct a 
prior predictive model~\cite{pdg2014,giunti} of our test statistic. When 
the systematic uncertainties are included, the significance of the 
observed signal, including the look-elsewhere effect, is reduced to 5.1$\sigma$.
For the analysis without the $\Delta R$ cut [Fig. 3(b)]
we obtain a significance including the systematic uncertainty and the look-elsewhere effect of 3.9$\sigma$. 

The stability of the result is checked by examining subsamples with
(i) different signs of the $\pi^{\pm}$ meson, (ii) different ranges of 
the azimuth and rapidity,
(iii) the distance between the $B_s^0$ vertex 
and the primary vertex changed  to five standard deviations, (iv) different $B_s^0$ mass
windows (1.7$\sigma$, 1.5$\sigma$, 1.2$\sigma$), (v) different $B_s^0 \pi^{\pm}$ momentum
intervals ($p_T >$ 9 GeV/$c$, $p_T >$ 12 GeV/$c$), and (vi) different
cone cuts ($\Delta  R <$ 0.2, $\Delta R <$ 0.15). Taking into 
account the  efficiencies of these cuts, no unexpected
behaviors are observed in these tests.

The invariant mass spectra of $B_s^0$ candidates and charged tracks
with kaon or proton mass hypotheses, are checked and 
no resonantlike enhancements in these distributions are found.

We measure the  ratio $\rho$ of the yield of the new state  $X(5568)$ to
the yield of the  $B_s^0$ meson in  two kinematic ranges,
10 $< p_T (B_s^0) <$ 15 GeV/$c$ and 15 $< p_T (B_s^0) <$ 30 GeV/$c$,
by repeating the $m(B_s^0 \pi)$ fits with  free mass and width  parameters
for the  $X(5568)$ signal~\cite{epaps}.
The results for $\rho$ are  (9.1 $\pm$ 2.6 $\pm$ 1.6)$\%$ and
(8.2 $\pm$ 2.7 $\pm$ 1.6)$\%$, respectively, with an average of (8.6 $\pm$ 1.9 $\pm$ 1.4)$\%$.
The systematic uncertainties due 
to $B_s^0$ reconstruction efficiency cancel out in the ratio.
The combined factor  of the soft pion kinematic acceptance, reconstruction efficiency, and selection
efficiency  is obtained from a simulated samples of events with a spinless particle
of mass equal to 5568 MeV/$c^2$ decaying to $B_s^0$ and a charged pion. 
The pion efficiency  increases with $p_T (B_s^0)$ from
(26.1 $\pm$ 3.2)$\%$ to (42.1 $\pm$ 6.5)$\%$ for the two $p_T (B_s^0)$
ranges. The systematic uncertainty due to a potential difference
of the soft pion reconstruction efficiency in MC calculations and data of $\pm$5$\%$
is accounted for  in systematics. Within uncertainties, the production ratio
$\rho$ does not depend on $p_T (B_s^0)$.

A possible interpretation of the observed structure is a four-quark state
made up of a diquark-antidiquark pair. With the $B_s^0 \pi^+$  produced
in an $S$-wave, its quantum numbers would be $J^{P}  = 0^{+}$. 
Thus, the state may be a heavy analog of the isotriplet scalar state
$a(980)$, with an $s$ quark replaced by a $b$ quark. Such open charm and
open bottom scalar mesons are predicted in Ref.~\cite{maiani}. On the 
other hand, the  state can decay through the chain 
$B_s^* \pi^{\pm}$, $B_s^* \to B_s^0 \gamma$, where the low-energy 
photon is not detected. In this case, the quantum numbers of this state
would be $J^{P}  = 1^{+}$, which would make it a counterpart to 
other heavy tetraquark candidates.
The mass of the new state would be shifted by addition of the nominal mass 
difference $m(B_s^*) - m(B_s^0)$, while its width would remain unchanged.
The large  difference  between the mass of this state and the sum of the $B_d$ and $K^\pm$ masses implies \cite{Karliner}  that $X(5568)$ is unlikely to be a molecular state composed of loosely bound $B_d$ and $K^\pm$ mesons.

In summary, a structure is seen in the $B_s^0 \pi^{\pm}$ invariant
mass spectrum  near threshold with a statistical significance,
 including the look-elsewhere effect, of 6.1$\sigma$. 
When the systematic uncertainties are included, the significance of the
signal is  5.1$\sigma$.
For the alternate analysis without the $\Delta R$ cut, we find the 
corresponding significance of 3.9$\sigma$.
This structure may be interpreted as a tetraquark state with four different
valence quark flavors, $b,s,u,d$.
The mass and natural width of the  $X(5568)$
state are 
$m = 5567.8 \pm  2.9  {\rm \thinspace (stat)} ^{+0.9}_{-1.9}  {\rm \thinspace (syst)}$  MeV/$c^2$
and 
$\Gamma = 21.9 \pm 6.4  {\rm \thinspace (stat)}   ^{+5.0}_{-2.5} {\rm \thinspace (syst)} $ MeV/$c^2$.

We thank E.~Gross and O.~Vittels for useful discussions.
We thank the staff at Fermilab and collaborating institutions,
and acknowledge support from the
Department of Energy and National Science Foundation (United States of America);
Alternative Energies and Atomic Energy Commission and
National Center for Scientific Research/National Institute of Nuclear and Particle Physics  (France);
Ministry of Education and Science of the Russian Federation, 
National Research Center ``Kurchatov Institute" of the Russian Federation, and 
Russian Foundation for Basic Research  (Russia);
National Council for the Development of Science and Technology and
Carlos Chagas Filho Foundation for the Support of Research in the State of Rio de Janeiro (Brazil);
Department of Atomic Energy and Department of Science and Technology (India);
Administrative Department of Science, Technology and Innovation (Colombia);
National Council of Science and Technology (Mexico);
National Research Foundation of Korea (Korea);
Foundation for Fundamental Research on Matter (Netherlands);
Science and Technology Facilities Council and The Royal Society (United Kingdom);
Ministry of Education, Youth and Sports (Czech Republic);
Bundesministerium f\"{u}r Bildung und Forschung (Federal Ministry of Education and Research) and 
Deutsche Forschungsgemeinschaft (German Research Foundation) (Germany);
Science Foundation Ireland (Ireland);
Swedish Research Council (Sweden);
China Academy of Sciences and National Natural Science Foundation of China (China);
and
Ministry of Education and Science of Ukraine (Ukraine).

\end{document}